\documentclass[aps,prb,twocolumn,preprintnumbers,amsmath,amssymb,colorlinks,showpacs,floatfix]{revtex4}

\usepackage{graphicx}
\usepackage{bm}
\usepackage{color}
\usepackage{natbib}
\usepackage{hyperref}
\usepackage{ulem}

\newcommand {\bcv} {BaCo$_2$V$_2$O$_8$}

\begin{document}

\title{Neutron diffraction investigation of the $H-T$ phase diagram\\ above the longitudinal incommensurate phase of \bcv}

\author{B. Grenier} 
\email[Corresponding author. Electronic address: ]{grenier@ill.fr}
\affiliation{INAC--SPSMS, Universit\'e Grenoble Alpes (UGA) and CEA, F-38000 Grenoble, France}

\author{V. Simonet, B. Canals, P. Lejay} 
\affiliation{Institut N\'eel, CNRS and UGA, F-38042 Grenoble, France}

\author{M. Klanj\v{s}ek} 
\affiliation{Jozef Stefan Institute, Jamova 39, SI-1000 Ljubljana, Slovenia}

\author{M. Horvati\'{c}, C.~Berthier} 
\affiliation{Laboratoire National des Champs Magn\'etiques Intenses, CNRS, UJF, UPS, and INSA, F-38000 Grenoble, France}


\begin{abstract}
The quasi-one-dimensional antiferromagnetic Ising-like compound \bcv\ has been shown to be describable by the Tomonaga-Luttinger liquid theory in its gapless phase induced by a magnetic field applied along the Ising axis. Above 3.9~T, this leads to an exotic field-induced low-temperature magnetic order, made of a longitudinal incommensurate spin-density wave, stabilized by weak interchain interactions. By single-crystal neutron diffraction we explore the destabilization of this phase at a higher magnetic field. We evidence a transition at around 8.5~T towards a more conventional magnetic structure with antiferromagnetic components in the plane perpendicular to the magnetic field. The phase diagram boundaries and the nature of this second field-induced phase are discussed with respect to previous results obtained by means of nuclear magnetic resonance and electron spin resonance, and in the framework of the simple model   based on the Tomonaga-Luttinger liquid theory, which obviously has to be refined in this complex system.
\end{abstract}

\maketitle


\section{introduction}

The universal Tomonaga-Luttinger liquid (TLL) theory describes the behavior of interacting fermions in one dimension (1D) where the Fermi liquid theory breaks down. A difference with the latter, characterized by quasiparticle excitations, is that the TLL gives rise to low-energy density wave excitations.\cite{tomonaga1950,luttinger1963,haldane1981,giamarchi2004} It is relevant for 1D conductors, as well as for the spin-1/2 quantum chain compounds. In this context, \bcv\ has been proposed as a model system realizing a TLL under a magnetic field.\cite{okunishi2007} \bcv\ crystallizes in the body-centered tetragonal $I4_1/acd$ (No.~142) space group, with $a=12.444$~\AA\, and $c=8.415$~\AA.\cite{wichmann1986} There are 16 spin-$\frac{3}{2}$ Co$^{2+}$ ions in a unit cell, which form screw chains running along the $c$ axis, two of which are described by a $4_1$ axis and the other two by a $4_3$ axis (see Ref.~\onlinecite{canevet2013} for more details on the crystallographic structure). As Co$^{2+}$ ions are located in a slightly distorted octahedral environment, they can be described by highly anisotropic effective $S=1/2$ spins, with a $g$-factor roughly twice larger along the $c$ axis than perpendicular to it.\cite{abragam1951,kimura2006,he2005,he2006} The intrachain antiferromagnetic (AF) coupling between the Co$^{2+}$ $S=1/2$ spins is much stronger than the interchain one, so that the system can be described by the $XXZ$ chain Hamiltonian:
\begin{equation}
{\cal H } = J \sum_i [\epsilon \left( S_i^x S_{i+1}^x + S_i^y S_{i+1}^y \right)  + S_i^z S_{i+1}^z]
\label{eq1}
\end{equation}
where $J$ is the intrachain AF interaction $(J>0)$ and $\epsilon$ is the anisotropy parameter. \bcv\ with $\epsilon < 1$ realizes an Ising-like case with the Ising axis aligned along the crystal $c$ axis.\cite{kimura2006,kimura2007,kimura2008a} The small but finite interchain coupling $J'$ leads to a long-range AF order below the N\'eel temperature $T_N\approx 5.5$~K, as first evidenced by He {\it et al.}.\cite{he2005} The zero-field magnetic structure [longitudinal antiferromagnetic (LAF) phase] was shown by neutron diffraction to consist of AF chains with the magnetic moments along the $c$ axis and exhibiting frustrated interchain couplings.\cite{kawasaki2010,canevet2013} In the TLL theory, the antiferromagnetic spin-spin correlation functions parallel (longitudinal) and perpendicular (transverse) to the applied magnetic field are of different nature: the former is incommensurate, with sinusoidally modulated correlations, while the latter is staggered (i.e.,'$+\,-\,+\,-$' correlations).\cite{haldane1980} They also have different field dependencies depending on the anisotropy. The exponent for the decay of the correlation functions with the distance between the spins is $\eta$ for the transverse fluctuations and $1/\eta$ for the longitudinal ones, where $\eta$ is termed the TLL exponent. \cite{okunishi2007} Then, whereas in the Heisenberg case the transverse fluctuations dominate in the gapless region ($\eta$ always $<1$ as soon as a magnetic field is applied), a more complicated scenario, departing from the classical picture, is proposed  in the Ising-like case: the longitudinal fluctuations should dominate only at low field ($\eta>1$), before being destabilized in favor of the transverse ones at higher field ($\eta<1$).\cite{okunishi2007} Applying a magnetic field along the Ising $c$ axis direction in \bcv\ closes the anisotropy gap at a critical field $H_c \sim 3.9$~T where a first transition occurs, below 1.8~K, from the zero-field N\'eel order to a new ordered phase that has inherited the longitudinal  incommensurate character from the parent TLL phase. The magnetic arrangement in this phase was determined by neutron diffraction as an incommensurate longitudinal spin density-wave (LSDW).\cite{kimura2008b,canevet2013} The modulation wave number $2k_F=2\pi(1/2-\left<S^z \right>)$ varies with the $z$ component of the field-induced spin $\left< S^z \right>$ due to the applied magnetic field, as expected in the TLL theory.\cite{haldane1980} The transition from this exotic ordered phase towards the transverse AF order is expected at a higher field $H^*$ before the saturated ferromagnetic (FM) state is reached at $H_s$. This magnetic-field region was probed both by electron spin resonance (ESR) above 1.3~K \cite{kimura2007} and by nuclear magnetic resonance (NMR) at lower temperature.\cite{klanjsek2015} The NMR experiment discovered and mapped two new low-temperature ordered phases between $8.6$~T and $H_s$, the transition between them being at $19$~T.\cite{klanjsek2015} This transition field coincides with the field where the anomaly of the high-field magnetization and the softening of one of the ESR modes were previously observed.\cite{kimura2007} Whereas a model based on weakly-coupled spin chains treated as TLLs\cite{okunishi2007} accounts remarkably well for the \bcv\ phase diagram up to 8.6~T, two additional ordered phases, not predicted in this model and whose exact nature is difficult to ascertain by   NMR alone, call for further investigation of this field region.

In this article, we present the results of single-crystal neutron diffraction under a magnetic field up to 12 T, applied along the Ising $c$ axis of \bcv, allowing us to identify the ordered magnetic phase above $8.6$~T, previously discovered by NMR,\cite{klanjsek2015} that borders on the LSDW phase. We find a transverse AF order compatible with a flop of the zero-field magnetic structure, as expected in the TLL picture, although at a lower field than the calculated one.\cite{okunishi2007} We discuss the result in light of the recent NMR investigation of this ordered phase.

\section{Results}

\begin{figure}
\begin{center}
\includegraphics[width=8.5 cm]{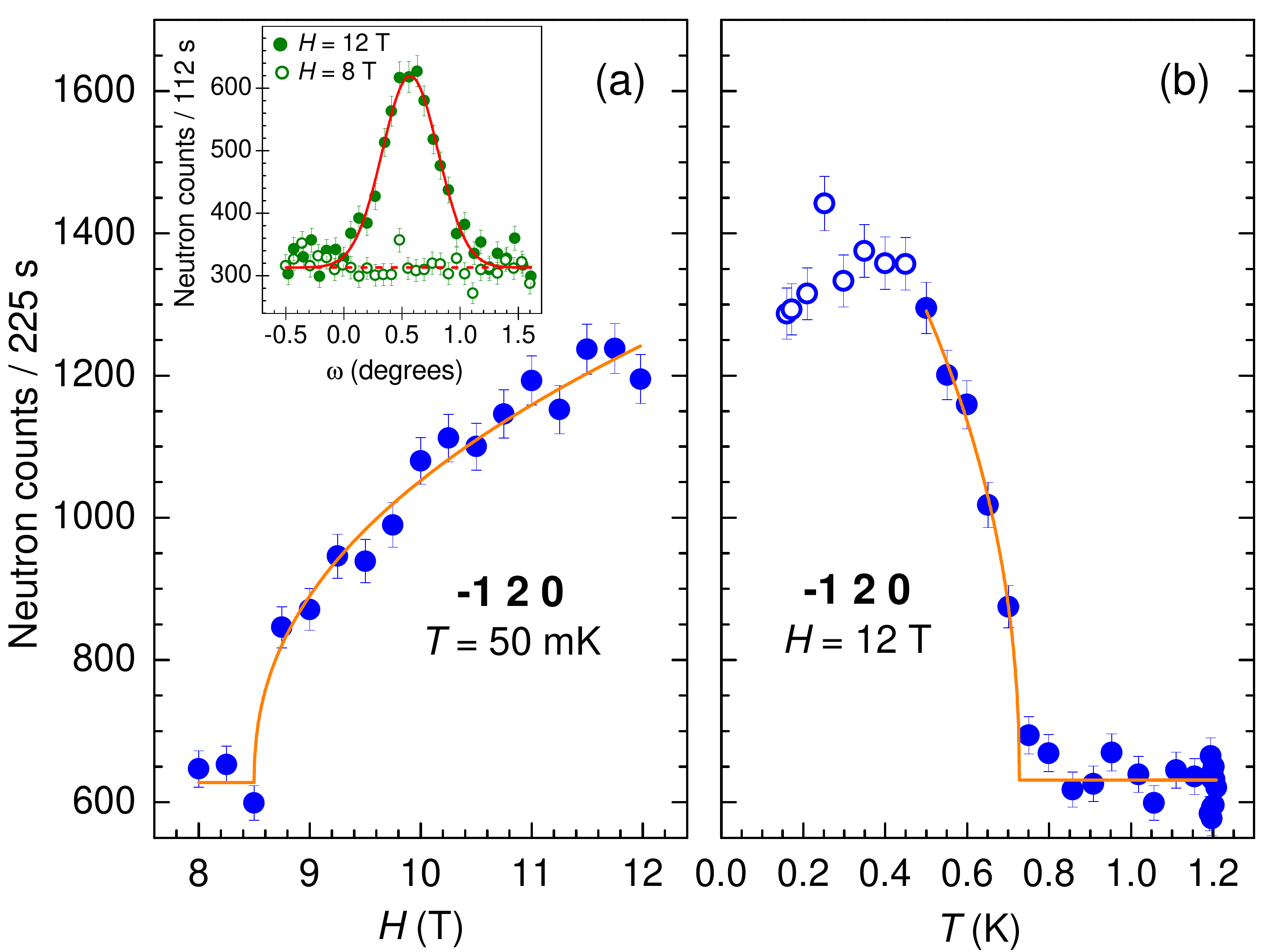}
\caption{(Color online) Field dependence at $T=50$~mK (a) and temperature dependence at $H=12$~T (b) of the magnetic signal: neutron counts on top of the $\bar{1}\,2\,0$ magnetic reflection of the high-field phase (blue circles). The orange lines are phenomenological fits to power laws allowing to extract the critical field $H^*=8.5$~T at 50~mK and critical temperature $T^*=0.7$~K at 12~T. The inset of panel (a) shows a rocking curve of the $\bar{1}\,2\,0$ reflection at $H=12$~T (fitted to a gaussian function) and at $H=8$~T where it is absent.}
\label{varH&T}
\end{center}
\end{figure}

A \bcv\ single-crystal was grown at Institut N\'eel by the floating zone method,\cite{lejay2011} from which a cubic-shaped crystal with about 3~mm long edges parallel to the crystallographic axes was cut. The neutron diffraction experiment was performed on the CEA--CRG D23 single-crystal two-axis diffractometer with a lifting arm detector at the Institut Laue Langevin high-flux reactor. The oriented sample was mounted on a dilution insert and placed in an Oxford vertical field 12 T cryomagnet, to have $H \parallel c$. Two different wavelengths were used, 2.38~\AA\ from a pyrolytic graphite monochromator and 1.28~\AA\ from a copper monochromator, in order to gain in flux and to access a larger volume of the reciprocal space, respectively. $hk0$ reflections (up to $h$ or $k=8$ with $2\theta^{\rm max}=120^\circ$) could be measured at the first wavelength, while $hk0$ and $hk1$ reflections (up to $h$ or $k=14$, restricting ourselves to $2\theta^{\rm max}=100^\circ$) could be measured at the smaller wavelength with an about three times smaller flux.

The high-field boundary of the LSDW phase, towards the second field-induced phase, was probed previously by single-crystal neutron diffraction \cite{canevet2013} and found to range from 9.25~T at 50~mK to 8.75~T at 480~mK. Above this temperature, the paramagnetic state is reached. Let us recall that the propagation vector of the LSDW phase is ${\bf k}_{LSDW}=(1,0,\xi)$ with $\xi$ the incommensurate modulation varying between 0.035 and 0.31~r.l.u. when the magnetic field, applied along the Ising $c$ axis, varies between the two critical fields $H_c \sim 3.9$~T and $H_p \sim 9.25$~T.\cite{canevet2013} We checked at 50 mK that the magnetic incommensurate satellite peaks associated with the LSDW phase indeed disappear at $H_p \sim 9.25$~T. Above this field, we looked for new magnetic Bragg peaks by scanning a portion of the reciprocal space. We found additional signal indexed by a propagation vector ${\bf k}_{AF} = (1, 0, 0)$, identical to the one of the zero-field N\'eel phase.\cite{canevet2013} This indicates a high-field commensurate AF structure [labelled transverse antiferromagnetic (TAF) as justified later on] with pure magnetic peaks appearing on $(h+1,k,l)$ positions with $h+k+l$ even, because of the body-centered lattice. Note that the intensity of the magnetic peaks in the high-field phase is two orders of magnitude smaller at 12~T than in the zero-field phase.

\begin{figure}
\begin{center}
\includegraphics[width=8.5 cm]{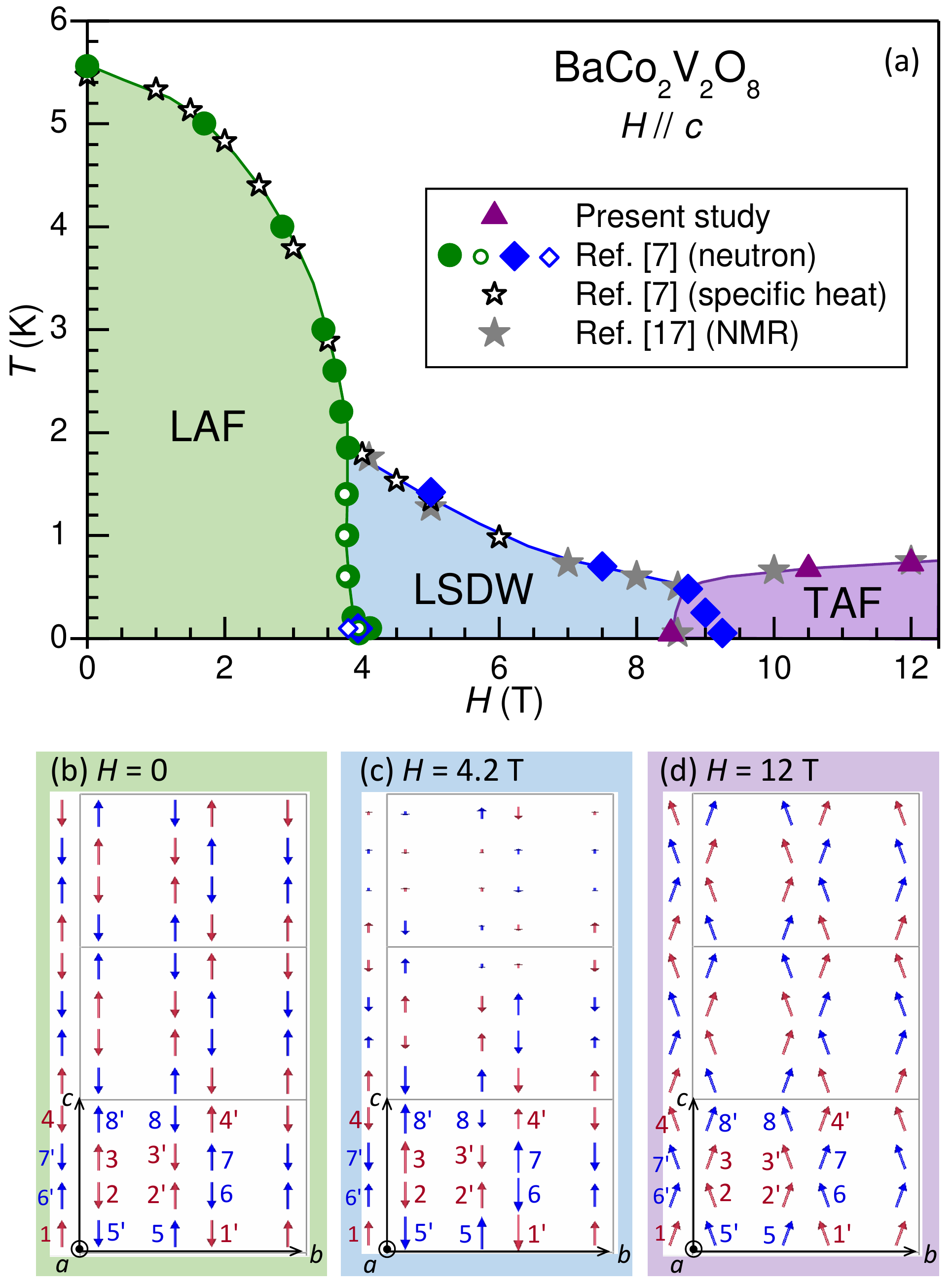}
\caption{(Color online) (a) Magnetic $H-T$ phase diagram of \bcv\ obtained from previous specific heat and neutron diffraction data,\cite{canevet2013} and completed at higher field by the present neutron diffraction study, in excellent agreement with the NMR data from Ref.~\onlinecite{klanjsek2015}. (b--d) Projection in the $(b,c)$ plane of one of the two domains of the magnetic structure corresponding to the LAF (b), LSDW (c), and TAF (d) phases. The TAF phase is canted due to the field-induced FM component, whereas this FM component is null at $H=0$ and not yet developed in the LSDW phase at 4.2~T, very close to the critical field. In the three panels, the $4_1$ and $4_3$ chains are in red and blue respectively. The labelling of the Co$^{2+}$ ions is explained in Fig. \ref{MagDom}. In panels (c) and (d), the scale for the length of the magnetic moments is respectively twice and three times that in panel (b).}
\label{DiagHT}
\end{center}
\end{figure}

In order to determine the boundaries of this phase in field and temperature, we followed the magnetic Bragg peak $\bar{1}\,2\,0$ when varying the magnetic field at $T=50$~mK and when varying the temperature at $H=10.5$~T (not shown) and $H=12$~T [see Figs.~\ref{varH&T}(a,b)]. This allowed us to complete the $H-T$ phase diagram of Ref.~\onlinecite{canevet2013} with this additional TAF ordered phase at higher fields [see purple triangles in Fig.~\ref{DiagHT}(a)]. On this phase diagram, the data points obtained from NMR\cite{klanjsek2015} are also shown (grey stars), evidencing a perfect agreement with our data for the LSDW and TAF phase boundaries. Interestingly, a coexistence of the LSDW and the TAF phases has been observed in our neutron data over a nearly 1~T wide region: the lower critical field of the TAF phase is at $H^* \simeq 8.5$~T, while the upper one of the LSDW phase is at $H_p \sim 9.25$~T. The LSDW--TAF phase transition is thus of first-order, like the LAF--LSDW one,\cite{canevet2013,suzuki2009} and contrary to the three transitions between each of these three ordered phases and the paramagnetic phase, which are all of second-order.
  
For the determination of the TAF magnetic structure, 56 different nuclear and 24 different magnetic reflections were collected at $H=12$~T and $T=50$~mK at $\lambda = 2.38$~\AA. Then, a few of these measurements were repeated at $\lambda = 1.28$~\AA, and additional reflections [not accessible at the previous wavelength] were added yielding in total (for both wavelengths) 142 different nuclear reflections and 32 different magnetic ones (reducing to 60 and 17 independent ones, respectively). The nuclear structure was first refined, using the two data sets and refining one scale factor for each wavelength, in addition to the 8 refinable $x,y,z$ coordinates and 4 $B_{iso}$ isotropic Debye-Waller factors of the Ba, Co, O1 and O2 atoms, and to the 4 extinction parameters.\cite{extinction,absorption} The nuclear structure was found to be identical to the one in the LAF and LSDW phases, with agreement $R-$factors $R_F=7.5$ and  $3.1$\% at $\lambda = 2.38$ and 1.28~\AA, respectively. This provided us with all the necessary information for the refinement of the magnetic structure. 

\begin{figure}
\begin{center}
\includegraphics[width=7.5 cm]{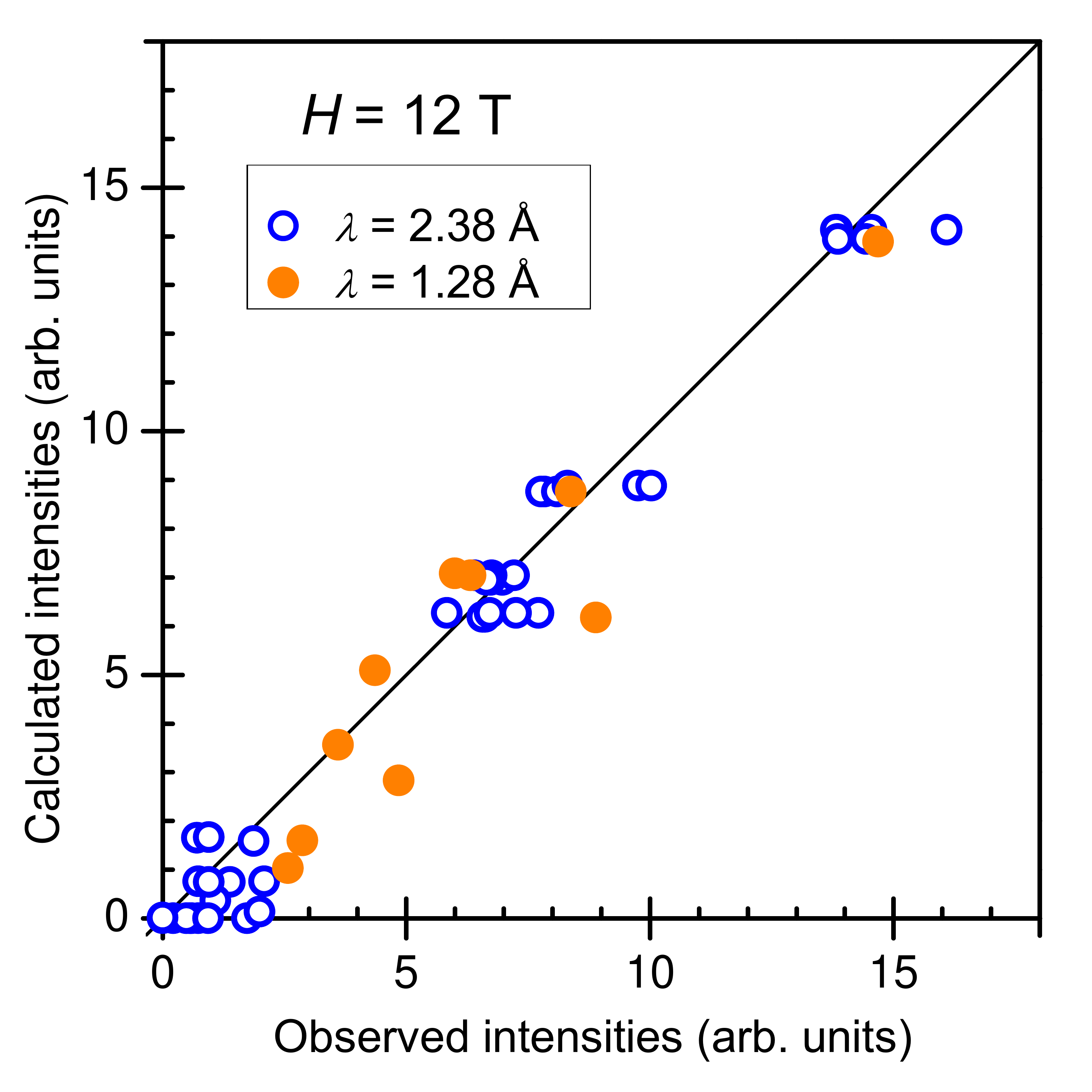}
\caption{(Color online) Graphical representation of the magnetic structure refinement in the high field TAF phase of \bcv\ at $T = 50$~mK and $H = 12$~T: calculated versus measured integrated intensities. The two sets of data collected at the two different wavelengths are rescaled to account for the difference in the neutron flux provided by the two monochromators, so that they can be presented on the same graph.}
\label{Fcalc_vs_Fobs}
\end{center}
\end{figure}

\begin{figure}
\begin{center}
\includegraphics[width=8.5 cm]{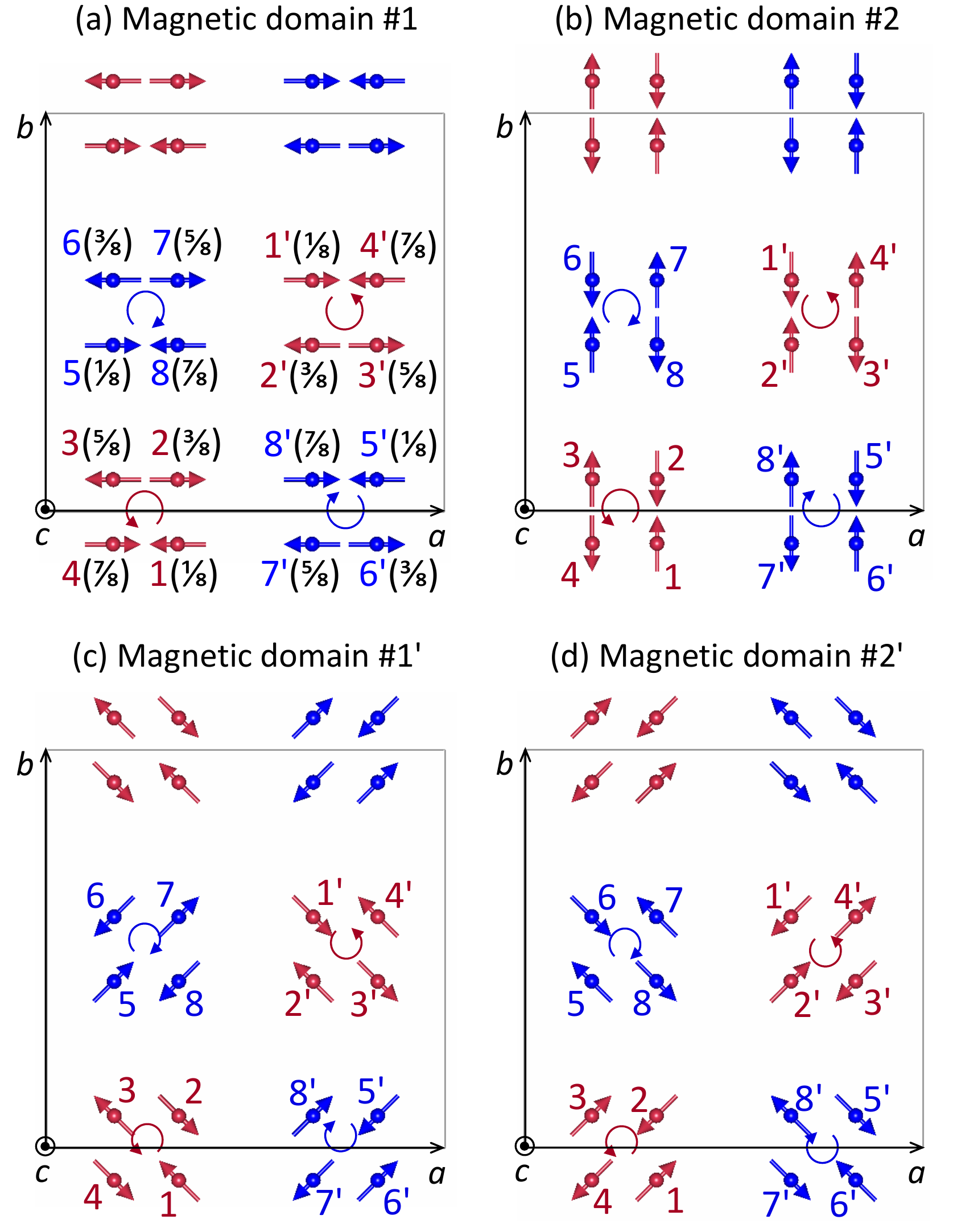}
\caption{(Color online) Projection in the $(a,b)$ plane of the two possible magnetic structures corresponding to the high-field TAF phase: (a) and (b) for the two magnetic domains of the collinear TAF structure with the AF component along the $\left< 100\right>$ directions; (c) and (d) for the two magnetic domains of the non-collinear TAF structure with the AF component along the $\left< 110\right>$ directions. The two magnetic domains are deduced from each other by application, e.g., of the $4_1$ or $4_3$ screw axis symmetry operator. The $z$ coordinate of the Co$^{2+}$ ions is given in parentheses in panel (a).}
\label{MagDom}
\end{center}
\end{figure}

The magnetic structure was then refined and the best result was obtained with agreement $R-$factors $R_F=10.8$ and 11.6\% ($R_{F^2 \omega}=9.55$ and 17.3\%) for $\lambda = 2.38$ and 1.28~\AA, respectively. The quality of the refinement can also be visualized in Fig.~\ref{Fcalc_vs_Fobs}. The high-field magnetic structure is characterized by a collinear arrangement of the AF component of the moments lying perpendicular to the applied field $H \parallel c$, so that this structure corresponds to a TAF ordering. Two domains are stabilized with the AF component along $a$ (domain \#1) or along $b$ (domain \#2). They are roughly equally populated with respective populations of $51.7 \pm 1.5$ and $48.3 \pm 1.5$\%. Indeed, the application of the magnetic field perpendicular to the $(a,b)$ plane should not favor one domain with respect to the other. The magnetic structure is plotted in Figs.~\ref{MagDom}(a,b) in projection onto the $(a,b)$ plane, for a better insight into the two magnetic domains and into the interchain couplings. Similarly to the zero-field LAF structure, this structure is compatible with an AF coupling along the chains and between two chains of the same type (either red $4_1$ chains or blue $4_3$ chains) in the diagonal $a \pm b$ directions. There is some remaining frustration between the neighboring chains of different types.
This can be seen for instance by the fact that one of the two equivalent pairs Co$_2$ -- Co$_8'$ and Co$_3$ -- Co$_{5}$ [through the $4_1$ screw axis located at $(\frac{1}{4}, 0, z)$] is ferromagnetic, while the other one is antiferromagnetic. The refined staggered component, $m_{TAF}=0.251(2)~\mu_B$/Co$^{2+}$ at $H=12$~T, is 9 times smaller than in the LAF phase at $H=0$.\cite{canevet2013} A field-induced FM component, parallel to the $c$ axis, superimposes on the AF one described above, resulting in a canted magnetic structure. This parallel component was determined in Ref.~\onlinecite{canevet2013}: at $H=12$~T, it is equal to 0.68~$\mu_B$/Co$^{2+}$, yielding a total moment amplitude of about 0.72~$\mu_B$/Co$^{2+}$.

Note that this magnetic structure cannot be univocally determined. Another AF magnetic structure can be obtained from the neutron data refinement at 12~T, with comparable agreement $R-$factors. This alternative structure corresponds to a non-collinear TAF structure. Again, there are two possible domains \#1' and \#2' that now give exactly the same intensities for all magnetic peaks, so that it is impossible to determine the domain populations, although they should be equal as for the collinear structure. The magnetic moments are along the $a-b$ and $a+b$ directions for the red and blue chains respectively in domain \#1' and reversed for the other domain [see Figs.~\ref{MagDom}(c,d)]. The AF component at 12~T is $m'_{TAF}=0.178(2)~\mu_B$/Co$^{2+}$, i.e., the previous value  divided by $\sqrt{2}$. The coupling is also AF along the chains and between two chains of the same type, but moments in two different types of chains are now at 90$^\circ$ from each other. This problem of indeterminacy in a tetragonal unit cell, between a collinear (moment $\parallel a$ or $\parallel b$) and a non-collinear (moment $\parallel a \pm b$) structure, with a $\sqrt{2}$ ratio in the moment amplitudes, is well known.\cite{brown2007} The use of polarized neutrons would be necessary to discriminate between both structures.

\section{Discussion}

\begin{figure}
\begin{center}
\includegraphics[width=8.5 cm]{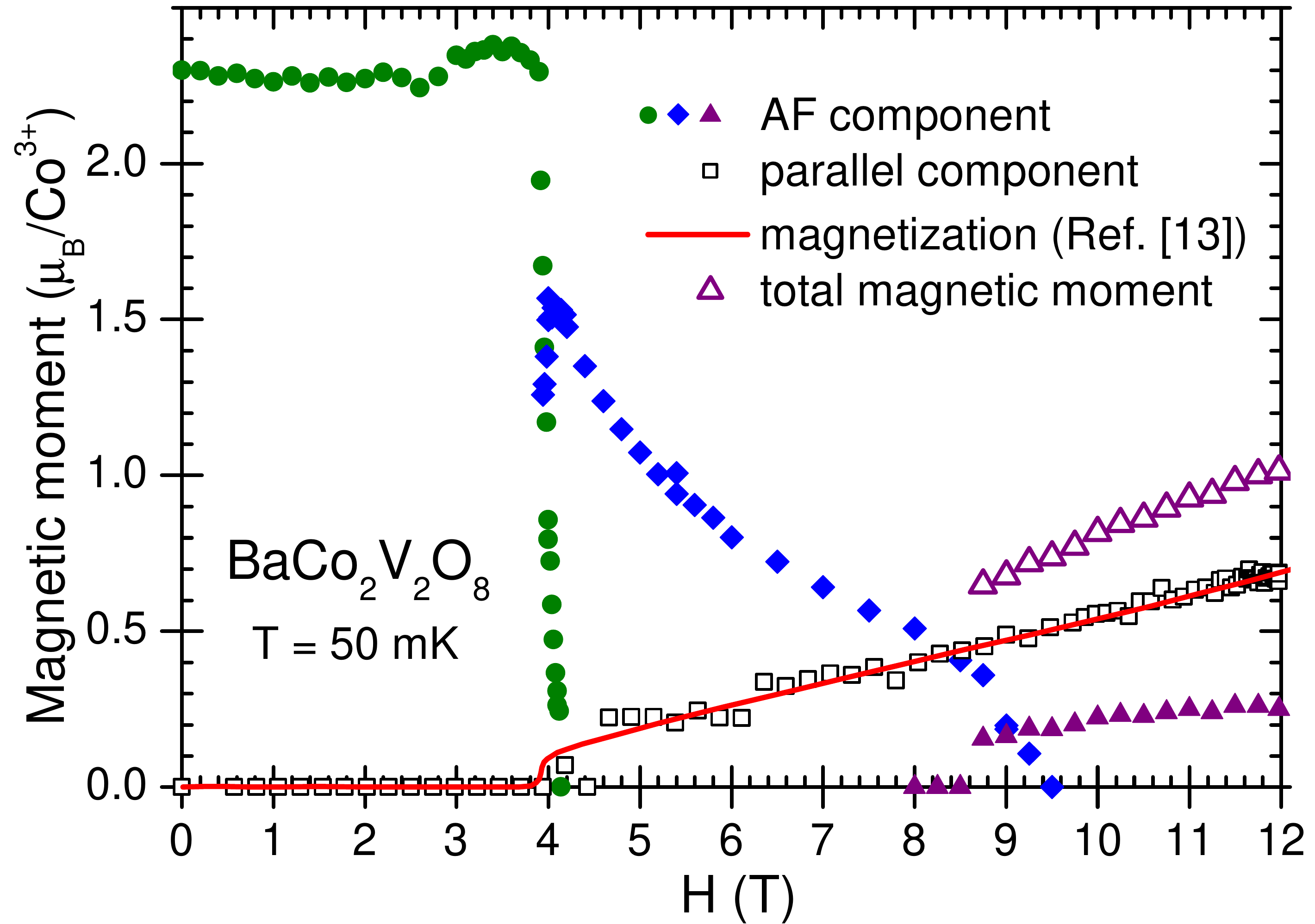}
\caption{(Color online) Field dependence of the various magnetic components determined at $T=50$~mK between 0 and 12~T by neutron diffraction in Ref.~\onlinecite{canevet2013} and in the present work (symbols) and from the magnetization measurement of Ref.~\onlinecite{kimura2008a} (solid line) after correction by a residual Van Vleck contribution ($m_{FM}$) as done in Ref.~\onlinecite{canevet2013}. The green circles and solid purple triangles correspond to the refined longitudinal ($m_{LAF}$) and transverse ($m_{TAF}$) antiferromagnetic components, respectively. The blue diamonds correspond to the refined amplitude $A$ of the sine wave function in the LSDW phase. The empty purple triangles give the total magnetic moment in the TAF phase $m_{tot}=\sqrt{m_{TAF}^2+m_{FM}^2}$.}
\label{IvsH}
\end{center}
\end{figure}

The present single-crystal neutron diffraction study under high magnetic fields allows us to  determine the phase diagram of \bcv\ up to $12$~T [see Fig.~\ref{DiagHT}(a)]. Figures~\ref{DiagHT}(b--d) display the projection, perpendicular to the $a$ axis, of each magnetic structure corresponding to the successive ordered phases stabilized with increasing field. The two low-field magnetic structures, LAF and LSDW, have their magnetic moments parallel to the applied magnetic field and are thus longitudinal. The high-field structure TAF is transverse, that is, with the antiferromagnetic component pointing in the plane perpendicular to the applied field, and appears as the flopped structure of the zero-field LAF one. Note that at the LSDW--TAF transition, the incommensurate modulation $\xi$ becomes close to the commensurate 1/3 value [see Fig.~9 of Ref.~\onlinecite{canevet2013}]. It is interesting to compare the evolution of the magnetic moment in the successive ordered phases. Fig.~\ref{IvsH} shows the field dependence of the ordered AF component across the three phases, together with the field-induced FM component and with the total magnetic moment in the TAF phase. 
One can see a large and abrupt jump in the amplitude of the  AF ordered component at the LAF--LSDW phase transition and only a slight decrease of this component at the LSDW--TAF phase transition. Moreover, close to the latter transition, the amplitude $A$ of the sinusoidal spin wave in the LSDW phase becomes equal to the field-induced FM component, meaning that there are no longer negative spin values (the AF longitudinal component opposite to the magnetic field is compensated by the FM one). Still around the critical field of this transition, the average of the absolute value of the modulated component in the LSDW phase ($2A/\pi$)  has significantly decreased so that it is nearly the same as the value of the ordered AF component in the TAF phase, which is probably a condition for the flop of the antiferromagnetic component perpendicular to the magnetic field. 
This updated $H-T$ phase diagram of \bcv\ qualitatively agrees with the TLL picture, with a TAF phase following at higher field the incommensurate LSDW phase. However, several points remain unsettled. 

First, the exact direction of the staggered AF component in the $(a, b)$ plane of the TAF phase cannot be unambiguously determined by our single-crystal neutron results, which are compatible both with a collinear and a non-collinear AF arrangement. At first sight, the orientation of the magnetic moments within the chain and between the chains for both magnetic structures can be explained by isotropic exchange interactions including: (i) a main intrachain AF interaction, (ii) an AF diagonal interaction between chains of the same type and (iii) frustrated interactions between chains of different types, like for the zero-field LAF structure. Additional ingredients must be invoked to explain the selection of the moment directions. Kimura {\it et al.} \cite{kimura2013} have noticed that the octahedral environment is slightly tilted from the $c$ axis. Their analysis of the magnetization measured for a magnetic field applied in the $(a,b)$ plane allowed them to determine the local $g$-tensor. In addition to the strongest anisotropy axis at $\simeq 5^{\circ}$ from the $c$ axis, a secondary anisotropy axis is also present perpendicularly, both axes rotating by 90$^{\circ}$ around the $c$ axis from one Co$^{2+}$ to the next one along the chain, under the effect of the screw axis. Such a single-ion anisotropy would favor neither the collinear nor the non-collinear magnetic structure. We have also calculated the dipolar energy and it was found to be identical for both magnetic structures, indicating that it is not a relevant parameter either. The Dzyaloshinskii-Moryia interaction is allowed between some Co$^{2+}$ pairs that lack an inversion center. However, its influence is difficult to grasp due to the lack of symmetry-based constraints on the direction of the Dzyaloshinskii-Moryia vector. Last, a most probable crucial ingredient in the selection of the field-induced magnetic state is the weak orthorhombic distortion that has been evidenced by thermal expansion and magnetostriction measurements.\cite{niesen2013} This distortion should favor coincident structural and magnetic domains with magnetic moments along the $a$ or $b$ directions, thus selecting the collinear magnetic structure. 

There also seems to be a discrepancy between the present neutron study and recent NMR results\cite{klanjsek2015} concerning the nature of the high field-phase. Indeed, Klanj\v{s}ek {\it et al.}\cite{klanjsek2015} observe at 10~T an NMR spectrum consisting of a single prominent peak that changes to a weak U-shaped spectrum at long echo-decay times. This spectrum is interpreted as a "ferromagnetic incommensurate (FM IC) ordering", where the Co$^{2+}$ chains would be ferromagnetically coupled in the $a$ and $b$ directions, with a modulation of the magnetic moment amplitude both along the $c$ axis and in the $(a,b)$ plane, the latter having a very long wavelength. We have been looking carefully for additional magnetic superlattice peaks at incommensurate positions $(h+\alpha,k\pm \alpha,l\pm \xi)$ and $(h,k\pm \alpha,l\pm \xi)$, by varying continuously the incommensurate modulations $\alpha$ and $\xi$ in various Brillouin zones. 
In spite of an optimization of the experimental conditions to increase the flux and of a long counting time, no signal could be detected elsewhere than at the AF position $(h+1, k, l)$. The influence of a misalignment of the magnetic field with respect to the $c$ axis is unlikely to explain the different phases deduced from neutron diffraction and NMR,\cite{explication} since the sample orientation could be precisely determined in our diffraction experiment and the crystal $c$ axis was found at less than 1 degree from the vertical field direction. Moreover, as can be seen in Fig.~\ref{DiagHT}, the high-field phase boundaries determined by NMR and neutron diffraction perfectly coincide indicating that both techniques probe the same magnetic phase. It is worth noting that the field dependence of the TLL exponent $\eta$ describing the decay of the spin-spin correlation functions, as determined from the field-dependence of the spin-lattice relaxation rate $T_1^{-1}$ in NMR, crosses the value $\eta = 1$ close to 8.5~T.\cite{klanjsek2015} This should correspond to the change from longitudinal incommensurate to transverse staggered fluctuations at this field value, in perfect consistency with the present neutron diffraction study. The two sets of experiments can be reconciled first by noticing that the TAF phase observed in neutron diffraction is compatible with the central prominent peak in the NMR line shape. Second, essentially all the NMR conclusions are based on the analysis of $T_1^{-1}$ probing the spin fluctuations. Above $H^*$, $T_1^{-1}$ sees essentially the incommensurate spin fluctuations and is weakly sensitive to the staggered AF ones due to the symmetric positions of the Co$^{2+}$ neighbors with respect to the probed $^{51}$V nucleus. This leads to a scenario in which the ordered phase is indeed the TAF phase determined by neutron diffraction, but dressed with strong incommensurate fluctuations (seen via the $T_1^{-1}$ in NMR), in addition to the expected staggered AF fluctuations. One cannot exclude also the true coexistence between the FM IC order proposed in the light of the NMR observations and the TAF magnetic order determined from our neutron diffraction study. The neutron diffraction experiment would not be sensitive to the weak incommensurate signal, as compared to the main TAF signal, which would be compatible with the NMR spectrum shown in Fig.~2e of Ref.~\onlinecite{klanjsek2015}.

Last, the agreement between our experimental result and a simple model of \bcv\, where weakly coupled $S=1/2$ $XXZ$ chains on a simple cubic lattice are treated as TLLs,\cite{okunishi2007} is only qualitative. Using the parameters $\epsilon = 0.46$ and $J'/J=0.00138$ extracted from the fit of the magnetization measurements,\cite{kimura2008a} the LSDW--TAF transition was calculated to occur at a magnetic field $H^* \sim 15$~T,\cite{okunishi2007} i.e., higher than $H^* \sim 8.5$~T as observed in our experiment. This disagreement may be simply due to the fact that the values of $\epsilon$ and $J'/J$ used by Okunishi and Suzuki\cite{okunishi2007} to determine this critical field are not correct. Indeed, our recent inelastic neutron scattering work suggests slightly different values ($\epsilon = 0.56$, $J=4.8$~meV, $J'=0.2$~meV yielding $J'/J=0.04$).\cite{grenier2015,greniererr} However, note that such a large $J'$ would drastically increase the critical temperature, which seems incompatible with the experimental phase boundaries. One possibility to resolve this issue is that interchain couplings are rather complicated, so that the frustration along the $a$ and $b$ directions produces an effective interchain coupling much smaller than the individual couplings. In this sense, it should also be noted that Okunishi and Susuki\cite{okunishi2007} have derived the phase diagram of \bcv\ in a mean-field model which does not take into account the fluctuations that can destabilize the LSDW order more rapidly than predicted in this approach. Another possibility to resolve the issue was suggested by Klanj\v{s}ek {\it at al.} in Ref.\onlinecite{klanjsek2015} where the combination of incommensurate fluctuations in spin chains and zigzag-like interchain couplings was shown to lead to the strongly field-dependent effective interchain coupling. Such a coupling allowed them to account for an additional phase, absent in the simple TLL picture, which they detected between 19 and 22 T by NMR, before the saturated FM state is reached.\cite{klanjsek2015} Finally, the complexity of the \bcv\ system is not fully captured by the simple model introduced in Ref.~\onlinecite{okunishi2007}: neither the orthorhombic distortion nor the complex structural arrangement in screw chains, that produces a multi-axis anisotropy, are accounted for. This simple model is clearly insufficient to account for all the properties of \bcv\ and for its rich $H-T$ phase diagram.

\section{Conclusion}

In conclusion, our neutron diffraction study has allowed us to determine the $H-T$ phase diagram of \bcv\ in a longitudinal field ($H$ parallel to the Ising $c$ axis) between 0 and 12~T. After a longitudinal AF phase (below $H_c\sim 3.9$~T) and a longitudinal spin-density wave phase (between $H_c$ and 9.25~T), we have identified a canted transverse AF phase above $H^*\sim 8.5$~T. On the one hand, the transverse nature of this field-induced magnetic order is compatible with the TLL picture used by Okunishi and Suzuki to calculate the \bcv\ phase diagram, \cite{okunishi2007} but the calculated critical field ($H^*\sim 15$~T) is higher than the measured one. On the other hand, this critical field matches the one deduced from the high-field NMR studies.\cite{klanjsek2015} However, the magnetic fluctuations above $H^*$ seen in NMR are incommensurate whereas we have identified a staggered transverse antiferromagnetic ordered phase by neutron diffraction. This may reflect an influence of coexisting longitudinal and transverse magnetic fluctuations in this range of magnetic fields. A calculation of the $H-T$ phase diagram using a more realistic model taking into account the complexity of \bcv\ is probably crucial to fully capture its rich physics.


We would like to thank E. Ressouche and S. Petit for fruitful discussions, and P. Fouilloux for his technical support. This work was partly supported by the French ANR project NEMSICOM.



\begin{references}

\bibitem{tomonaga1950} S.~Tomonaga, Prog. Theor. Phys. {\bf 5}, 544 (1950).

\bibitem{luttinger1963} J.~M.~Luttinger, J. Math. Phys. {\bf 4}, 1154 (1963).

\bibitem{haldane1981} F.~D.~M.~Haldane, J. Phys. C {\bf 14}, 2585 (1981).

\bibitem{giamarchi2004} T.~Giamarchi, Quantum Physics in One Dimension (Oxford University Press, Oxford, 2004) and references therein.

\bibitem{okunishi2007} K.~Okunishi and T.~Suzuki, Phys. Rev. B. {\bf 76}, 224411 (2007).

\bibitem{wichmann1986} R.~Wichmann and Hk.~M$\ddot{\rm u}$ller-Buschbaum, Z.~Anorg. Allg. Chem. {\bf 532}, 153 (1986).

\bibitem{canevet2013} E.~Can\'evet, B.~Grenier, M.~Klanj\v{s}ek, C.~Berthier, M.~Horvati\'{c}, V.~Simonet, and P.~Lejay, Phys. Rev. B {\bf 87}, 054408 (2013), and references therein.

\bibitem{abragam1951} A.~Abragam and M.~H.~L.~Pryce, Proc. R. Soc. Lond. A {\bf 206}, 173 (1951).

\bibitem{kimura2006} S.~Kimura, H.~Yashiro, M.~Hagiwara, K.~Okunishi, K.~Kindo, Z.~He, T.~Taniyama, and M.~Itoh in {\it Yamada Conference LX on Research in High Magnetic Fields, Sendai, Japan, 2006} [J. Phys.: Conf. Ser. {\bf 51}, 99 (2006)].

\bibitem{he2005} Z.~He, D.~Fu, T.~Ky\^omen, T.~Taniyama, and M.~Itoh, Chem. Mater. {\bf 17}, 2924 (2005).

\bibitem{he2006} Z.~He, T.~Taniyama, and M.~Itoh, Appl. Phys. Lett. {\bf 88}, 132504 (2006).

\bibitem{kimura2007} S.~Kimura, H.~Yashiro, K.~Okunishi, M.~Hagiwara, Z.~He, K.~Kindo, T.~Taniyama, and M.~Itoh, Phys. Rev. Lett. {\bf 99}, 087602 (2007).

\bibitem{kimura2008a} S.~Kimura, T.~Takeuchi, K.~Okunishi, M.~Hagiwara, Z.~He, K.~Kindo, T.~Taniyama, and M.~Itoh, Phys. Rev. Lett. {\bf 100}, 057202 (2008).

\bibitem{kawasaki2010} Y.~Kawasaki, J.~L.~Gavilano, L.~Keller, J.~Schefer, N.~B.~Christensen, A.~Amato, T.~Ohno, Y.~Kishimoto, Z.~He, Y.~Ueda, and M.~Itoh, Phys. Rev. {\bf B 83}, 064421 (2011).

\bibitem{haldane1980} F.~D.~M.~Haldane, Phys. Rev. Lett. {\bf 45}, 1358 (1980).

\bibitem{kimura2008b} S.~Kimura, M.~Matsuda, T.~Masuda, S.~Hondo, K.~Kaneko, N.~Metoki, M.~Hagiwara, T.~Takeuchi, K.~Okunishi, Z.~He, K.~Kindo, T.~Taniyama, and M.~Itoh, Phys. Rev. Lett. {\bf 101}, 207201 (2008).

\bibitem{klanjsek2015} M.~Klanj\v{s}ek, M.~Horvati\'c, S.~Kr\"amer, S.~Mukhopadhyay, H.~Mayaffre, C.~Berthier, E.~Can\'evet, B.~Grenier, P.~Lejay, and E.~Orignac, Phys. Rev. B, in press.

\bibitem{lejay2011} P.~Lejay, E.~Can\'evet, S.~K.~Srivastava, B.~Grenier, M.~Klanj\v{s}ek, and C.~Berthier, J. Cryst. Growth {\bf 317}, 128 (2011).

\bibitem{suzuki2009} T.~Suzuki, N.~Kawashima, K.~Okunishi, Journal of Physics: Conference Series {\bf 150}, 042197 (2009).

\bibitem{extinction} see Refs. 21 and 22 in Ref.~\onlinecite{canevet2013} for more details on the refinement.

\bibitem{absorption} Note than correcting the data by the absorption is rather useless due to the cubic shape of the crystals (the transmission coefficients from one Bragg reflection to another varies by less than 1\%).

\bibitem{brown2007} P.~J.~Brown, Collection SFN {\bf 7}, 123 (2007), EDP Sciences, Les Ulis, http://dx.doi.org/10.1051/sfn:2007020.

\bibitem{kimura2013} S.~Kimura, K.~Okunishi, M.~Hagiwara, K.~Kindo, Z.~He, T.~Taniyama, M.~Itoh, K.~Koyama, and K.~Watanabe, J. Phys. Soc. Japan. {\bf 82}, 033706 (2013).

\bibitem{niesen2013} S.~K.~Niesen, G.~Kolland, M.~Seher, O.~Breunig, M.~Valldor, M.~Braden, B.~Grenier, and T.~Lorenz, Phys. Rev. B {\bf 87}, 224413 (2013).

\bibitem{explication} Klanj\v{s}ek {\it et al.}\cite{klanjsek2015} have noticed that the FM IC phase they observed was very sensitive to the field direction: with only a few degrees of misalignment between $H$ and the $c$ axis, the phase boundary gets significantly reduced in temperature and eventually disappears at higher fields.

\bibitem{grenier2015} B.~Grenier, S.~Petit, V.~Simonet, E.~Can\'evet, L.-P.~Regnault, S.~Raymond, B.~Canals, C.~Berthier, and P.~Lejay, Phys. Rev. Lett. {\bf 114}, 017201 (2015).

\bibitem{greniererr} B.~Grenier, S.~Petit, V.~Simonet, E.~Can\'evet, L.-P.~Regnault, S.~Raymond, B.~Canals, C.~Berthier, and P.~Lejay, erratum of Phys. Rev. Lett. {\bf 114}, 017201 (2015), submitted.

\end{references}
\end{document}